\documentstyle[12pt]{article}
\begin{document}
\begin{titlepage}
\pagestyle{empty}
\baselineskip=21pt
\rightline{OSU-TA-2/95}
\rightline{UMN-TH-1325/95}
\rightline{hep-ph/yymmddd}
\rightline{February 1995}
\vskip .2in
\begin{center}
{\large{\bf A New Look At Neutrino Limits From \\
 Big Bang Nucleosynthesis}} \end{center}
\vskip .1in
\begin{center}
Keith A. Olive

{\it School of Physics and Astronomy, University of Minnesota}

{\it Minneapolis, MN 55455, USA}

and

Gary Steigman

{\it Department of Physics, The Ohio State University}

{\it Columbus, OH 43210, USA}

\vskip .1in

\end{center}
\vskip .2in
\def\li#1{\hbox{${}^{#1}$Li}}
\def\he#1{\hbox{$^{#1}{\rm He}$}}
\centerline{ {\bf Abstract} }
\baselineskip=18pt
We take a fresh look at the limits on the number of neutrino flavors derived
from big bang nucleosynthesis.  In particular, recent measurements of the
\he4 abundance enable one to estimate the primordial \he4 mass fraction at
$Y_p = 0.232 \pm .003(stat) \pm .005(syst)$.  For a baryon to photon ratio,
 $\eta$,
consistent with the other light elements, this leads to a best fit
for the number of neutrino flavors $N_\nu < 3$ (the precise number
depends on $\eta$) indicating a very strong upper limit to $N_\nu$.  Here, we
derive new upper limits on $N_\nu$, paying special attention to the
fact that the best estimate may lie in an unphysical region ($N_\nu < 3$ if all
three neutrino flavors are light or massless;
the lower bound to $N_\nu$ may even be as low as 2, if the small window
for a $\nu_\tau$ mass
is exploited).  Our resulting upper limits therefore depend on
whether $N_\nu \ge 2$ or 3 is assumed.  We also explore the sensitivity
of our results to the adopted value of $\eta$ and the assumed systematic
errors in $Y_p$.

\noindent
\end{titlepage}
\baselineskip=18pt

\def\la{{{\lower 5pt\hbox{$<$}} \atop {\raise 5pt\hbox{$\sim$}}}~}
\def\ga{{{\lower 2pt\hbox{$>$}} \atop {\raise 1pt\hbox{$\sim$}}}~}
\def\li#1{\hbox{${}^{#1}$Li}}
\def\he#1{\hbox{$^{#1}{\rm He}$}}
\def\N{$N_\nu$~}
\def\beq{\begin{equation}}
\def\eeq{\end{equation}}

Among the strongest cosmological constraints on particle physics
models are those derived from the consistency of standard big bang
nucleosynthesis (SBBN) and, of particular importance are the constraints
determined from the consistency of the predicted and observed primordial
 $^4$He mass fraction, $Y_p$.
Most notable among these constraints is the limit on the number of
neutrino flavors \cite{ssg} which can be translated into a host of other
limits on particle physics properties.
In the SBBN \cite{wssok}, the abundances are primarily sensitive to a
single parameter, the baryon-to-photon ratio, $\eta$. Consistency between
 the predictions of SBBN and the observational determinations of the
light element abundances restricts $\eta$ to a narrow range between
$2.8 \times 10^{-10}$ and $4 \times 10^{-10}$ \cite {wssok}. In this range
 the calculated
$^4$He  mass fraction lies in the range $Y_p = 0.240 - 0.245$ \cite{kern},
which is high when compared with the observationally
inferred best primordial value \cite{OSt},
\beq
Y_p = 0.232 \pm 0.003 \pm 0.005
\label{yost}
\eeq
where the errors are 1 $\sigma$ statistical and systematic errors
respectively. Indeed, consistency within these small errors
 allows for very little room for any enhancement in
primordial $^4$He.  This is the reason that SBBN leads to very tight
limits on the number of neutrino flavors \cite{ssg,OSt,kk}.

Indeed, based on our estimate for the primordial mass fraction of
\he4, we deduced a best fit for the number of neutrino flavors \cite{OSt},
\beq
N_\nu = 2.17 \pm 0.27 \pm 0.42
\eeq
In \cite{OSt} we used a $2 \sigma_{\rm stat} + \sigma_{\rm sys}$ upper
limit for
$Y^{\rm max}$  in order to test the consistency of the SBBN.
That procedure yields $N_\nu < 3.13$, a value similar
to, but somewhat weaker than the 95 \% CL limit $N_\nu < 3.04$ from
a Monte Carlo analysis \cite{kk}.  However, if there are indeed three
massless neutrinos, then our best fit ($N_\nu = 2.17$) is in fact in the
unphysical regime.  Limits derived from a probability distribution
centered in an
unphysical region of parameter space are known to give overly restrictive
bounds.\footnote{We thank Paul Langacker for bringing this issue to our
attention.} In this letter the limits on $N_\nu$ are rederived by
renormalizing the probability distribution to the physical portion of
parameter space.  We consider both $N_\nu \ge 3$ and $N_\nu \ge 2$
as criteria for the physical region.  The limits on $N_\nu$ are of
course weakened, even in the latter case ($N_\nu > 2$) where though the mean
 value is physical, much of the probability distribution still lies in
an unphysical region ($N_\nu < 2$).

The best approach to a BBN bound on $N_\nu$ is to fit simultaneously both
$\eta$ and $N_\nu$ to the inferred primordial abundances of all the light
elements \cite{new}.  Here,
instead, we simply explore these new limits to
 $N_\nu$ as a function of $\eta$, as
well as of the assumed systematic uncertainty.
  However, treatment of the systematic
uncertainty can not always be done rigorously.
  We therefore study the implications of
two different assumptions.  We will assume either
that they are Gaussian, in which case they will add in
quadrature with the statistical errors or, that they
are described by a top-hat distribution which we convolve with the Gaussian
statistical errors (this is one of the cases being considered in \cite{kern}).
  We also discuss the consequences of having several
sources of systematic errors on our estimates of $N_\nu$ and $Y^{\rm max}$.

As a prelude to our statistical analysis, it is useful to discuss first
 the relationship between $N_\nu$ and the \he4 abundance \cite{ssg}.
The $^4$He abundance is primarily determined by the neutron-to-proton
ratio just prior to nucleosynthesis or, more accurately, prior to the
freeze-out
of  the weak interaction rates at a temperature $T_f \sim 1$ MeV. The final
\he4 abundance is in fact quite sensitive to the freeze-out temperature which
is determined by the competition between the weak interaction rates and the
expansion rate of the Universe.  As the uncertainty in the neutron mean-life
is now very small, $\tau_n = 887 \pm 2$ s \cite{rpp}, the helium
abundance depends primarily on $\eta$ which determines the onset of
nucleosynthesis and
\N, which can be used to characterize the expansion rate of the Universe at
$T \ga 1$ MeV, as measured by the (time dependent) Hubble parameter H,
\beq
H = \left( {8 \pi^2 G_N (N_* + {7 \over 4} N_\nu) \over 90} \right)^{1/2} T^2
\label{h}
\eeq
where $N_* = 5.5$ provides the contribution from electrons and photons.
It is appropriate to emphasize that $N_\nu$ measures the contribution to the
energy density at the epoch of nucleosynthesis of neutrinos (massive or
massless) and any additional particles beyond those in the standard model.
In the standard model, \N = 3 (unless the $\tau$ neutrino has a mass in
excess of about 0.1 MeV) \cite{Kaw}.  Limits on additional
massless neutrino flavors
with standard model coupling to the $Z$ gauge boson are very tightly
constrained by LEP: \N $<$ 3.04 (2 $\sigma$) \cite{rpp}.  The nucleosynthesis
bounds however are most sensitive to {\em any} relativistic  particle species
present at that time. For example, an additional massless scalar degree of
freedom such as the majoron, would contribute 4/7 to \N.

The neutron-proton ``freeze-out" is determined by the competition between the
expansion rate (\ref{h}) and the weak interaction rate
 (proportional
to $T^5$). A larger value for \N increases the expansion rate leading to
earlier freeze-out at a higher neutron to proton ratio and, hence, yields
 a higher value for $Y_p$.
To understand the behavior of the calculated
 value of $Y_p$ as $\eta_{10}, \tau_n$
and $N_\nu$ are varied, it is useful to fit the SBBN results in the form
\beq
Y_p = A + B\ln \eta_{10} + C(\tau_n - 887)
+ D(N_\nu - 3)
\label{yp}
\eeq
As $\eta_{10}$ increases from 2 to 10, B decreases from 0.014 to 0.009.  For
all
$\eta$, $C = 2 \times 10^{-4}$, confirming that the small uncertainty in the
neutron lifetime contributes negligibly to the uncertainty in $Y_p$.  As
$N_\nu$
varies from 2 to 4, D decreases from 0.014 to 0.012. From eq. (\ref{yp}), it is
clear that to bound $N_\nu$ will require that $Y_p$ and $\eta_{10}$ and their
uncertainties be constrained.

In Yang et al. \cite{ytsso} and Walker et al. \cite{wssok},
we utilized solar system abundances of D and \he3 to place an upper bound
to the primordial abundance of D + \he3 and we derived a lower bound to
 $\eta_{10}~(> 2.8$).
 This bound is conservative in the sense that it
ignored \he3 production in low mass stars. Though models of galactic chemical
evolution using primordial abundances near this lower bound on $\eta$ may
show sufficient deuterium destruction \cite{vop},  when \he3 production is
included these models predict a large excess of
\he3
\cite{orstv} unless $\eta$ is large $(\ga 4)$ \cite{st}. Indeed, a recent
Monte Carlo analysis \cite{hsstw} found the larger value of
 $\eta_{10} \sim 6.6 \pm 1.4$
  as  a best fit even when \he3 production was ignored.  Thus, the challenge
  to BBN is clear since,
 even for $\eta_{10} \ge 2.8$ and $\tau_n \ge 885$, we
have $Y_p \ge 0.240$ for  \N = 3 .  As was stressed in \cite{OSt}, this is
 only consistent with the observational data if $\sigma(syst) = 0.005$ is
 added to the 2$\sigma_{\rm stat}$ upper bound of 0.238.

Recently, there have been reports of observations of deuterium in quasar
absorption
systems \cite{cs}  with a D/H abundance (which may be interpreted to be
the primordial one) corresponding to a value of
$\eta_{10} \sim 1.5$. Although such a low value for $\eta$ would be
consistent with the observed \he4 abundance, it would exacerbate the
problem due to the overproduction of \he3 \cite{s,vc}. We note that there
is also the
possibility that this observation can be interpreted as a H detection  in
which the absorber is displaced in velocity by 80 km s$^{-1}$ with respect to
the quasar \cite{s}. Such an interpretation is likely, given recent
reports \cite{t} that a D/H measurement along a different line of sight may
indicate a much lower value for D/H, corresponding to a much higher value for
$\eta_{10} (\ga 6$). If this latter is correct, it would appear to pose a
conflict between the BBN predictions and the
observational determinations of $Y_p$. At this point, however, the data on D/H
is too limited (and lacks consistency) to permit any firm conclusion. In our
subsequent discussion of limits on \N, we will, where appropriate, give
results for several values of $\eta_{10}$.

 Before returning to the problem of placing limits on \N, we first
discuss our treatment of errors used in establishing a $2\sigma$ (or $\sim 95
\%$ CL) upper limit.  The chief problem is the treatment of systematic errors.
In obtaining $Y^{\rm max} = 0.243$ \cite{OSt},
 we simply added $\sigma_{\rm sys} =
0.005$ to
the $2\sigma$ statistical upper limit.
 There are better alternatives.  For example,
it can be assumed (as is done by the particle data group \cite{rpp}) that
systematic errors are Gaussian distributed and add in quadrature to the
statistical errors.  In this case, the 2$\sigma$ upper limit to $Y_p$ is
0.244.  Or, it may be assumed that the systematic errors are described by a
top-hat distribution (ie. constant probability between $\pm
\sigma_{\rm sys}$ and zero otherwise). The
convolution of the Gaussian and top-hat distributions is just the difference
of error functions given by (unnormalized)
\beq
{\rm erf}\left( {x + \sigma_{sys} - \mu \over \sqrt{2} \sigma_{\rm stat}}
\right) - {\rm erf}\left( {x - \sigma_{sys} - \mu \over \sqrt{2} \sigma_{\rm
stat}}
\right)
\label{ytg}
\eeq
where $\mu = 0.232$ is the central value for $Y_p$. For $\sigma_{\rm stat} =
0.003$ and $\sigma_{\rm sys} = 0.005$, the distribution given in eq.
(\ref{ytg}) yields a 95 \% CL upper limit of 0.240 which is somewhat more
restrictive than the previous two estimates.

It is very difficult to estimate the size of the systematic errors.  Indeed,
in exploring possible sources of such errors it is important to consider
whether they may be correlated or
 uncorrelated among themselves \cite{kern,new}.  It has
recently been suggested that previous analyses have overlooked potential
sources of error and, therefore, that the true systematic uncertainty may
be considerably larger than our estimate of 0.005 \cite {sg,cst}.
Although we do agree that the size of the systematic error remains uncertain,
we consider that the recent analyses may be overly simplified.
To illustrate the problem, consider the contributions to the uncertainty from
corrections for ionization and collisional
 excitation \cite{kern,new}.  It is often assumed
that there could be some neutral helium in the zone where hydrogen is fully
ionized and, therefore, that the ionization correction will only increase
the inferred helium abundance \cite{sg,cst}.  However, this overlooks the
fact that very metal-poor stars will tend to be hotter than their higher
metallicity counterparts, resulting in a harder ionizing radiation field
in the very metal-poor HII regions.  In such a situation, the HeII zone may
extend beyond the HII zone and an appropriate ionization correction would
reduce the inferred He abundance.  Even worse, this possible effect would
then correlate with the metallicity of the HII region, perhaps reducing Y at
low metallicity and raising it for more metal-rich HII regions.  The same
considerations apply for the possible correction to account for collisional
excitation in helium.  This correction (decreasing the inferred helium
abundance) will be larger for hotter (metal-poor) regions and smaller for
cooler (metal-rich) ones.

Thus, since some of the sources of systematic errors may be correlated (or
anticorrelated) and, even their sign uncertain, a naive linear combination
of several error estimates \cite{sg,cst} must lead to an overestimate of the
true uncertainty.  To explore the effect of such possible errors on our
estimate of $Y^{\rm max}$, we consider
 some illustrative examples.   For example,
let us first assume that the systematic errors are Gaussian distributed.  Then
we simply add, in quadrature, the systematic error(s) to the statistical one.
Thus, as noted above, with our choice
 of 0.005 for the systematic error and 0.003 for
the statistical one, we find for the 95 \% CL upper limit on $Y$, $Y_{95}$ =
0.244.  For two uncorrelated systematic errors, each of size 0.005, we would
increase this to
$Y_{95}$ = 0.247. If, instead, the systematic error were 0.010, we would find
$Y_{95}$ =  0.253 ; note the relatively large difference if the systematic
errors are added in quadrature or, linearly.  In contrast, if we we treat the
systematic errors as uniform (``top-hat"), the probability distribution for $Y$
is given by eq. (\ref{ytg}), and for the three cases just considered we  find
$Y_{95}$ = 0.240, for a Gaussian statistical error of 0.003 and a top-hat
systematic error of 0.005; 0.242, for a Gaussian statistical error of
 0.003 and two top-hat
systematic errors of 0.005 each (in this case one must convolve a second
top hat with the distribution in eq. (\ref{ytg}) which is
a distribution given by a linear combination of four error functions and
four Gaussians); and, 0.244 for a
 Gaussian statistical error of 0.003 and a single
top-hat systematic error of 0.010.  These ``2 $\sigma$" upper limits are
considerably smaller than those from the double Gaussian approach and certainly
 much smaller than
the upper bound obtained by adding the systematic errors linearly.

We turn now to our computation of the upper limit on \N.
To begin, we take $Y_p$ and its uncertainties from eq. (\ref{yost}) and
compare to the central predicted value of $Y_p$ for each value of $\eta_{10}$
and $\tau_n$ = 887 sec. to find the best (i.e., central) value for \N.  For
example, as $\eta_{10}$ increases from 1 to 10, the central value of
\N decreases from 3.6 to 1.1; the best values of \N corresponding to
 $\eta_{10}$ = 2.8 and 4.0,
are 2.3 and 1.9 respectively.  If all three neutrinos were in fact
relativistic at the time of nucleosynthesis, then these latter values (and,
indeed,
all values for $\eta_{10} > 1.5$) are unphysical. For the central case of
$\eta_{10} = 3$, even the 2$\sigma$ statistical upper limit is unphysical.
When the systematic uncertainty is included, the upper bound goes above
3.0, but this is most certainly an overly restrictive upper bound based on
the data at hand.  Here we will use the Bayesian approach described in
\cite{rpp}.  Consider first, the double Gaussian treatment of errors
(ie., both statistical and systematic errors as Gaussians). In the
Gaussian distribution
\beq
f (x) = e^{{-(x - N_\nu)^2 \over 2 \sigma^2}}
\eeq
$\sigma^2 = \sigma_{\rm stat}^2 + \sigma_{\rm sys}^2$ and \N is the
(possibly unphysical) value determined for a given value of
$\eta_{10}$. Normally, this distribution would be normalized by integrating
over $x$ from $-\infty$ to $\infty$ and setting the result equal to unity.
The 95
\% CL limits on $x$ would
 correspond to the limits of integration of the normalized
distribution integrated to the value 0.95.  Instead, here we allow the
distribution to take non-zero values only over the physical region
($N_\nu \ge 3$). The distribution is now normalized by integrating from 3 to
$\infty$ and the 95 \% CL upper limit is the upper limit of integration
$x_{95}$  so that the renormalized distribution yields a value 0.95 when
integrated from 3 to $x_{95}$,
\beq
{\int^{x_{95}}_3 f(x) dx \over \int^{\infty}_3 f(x) dx} = 0.95
\label{int}
\eeq
  The 95\% CL upper limit on \N as a function
of $\eta_{10}$ is shown in Figure 1 by the curve labeled $N_\nu \ge 3$.
For $\eta_{10} > 2.8$, \N $< 3.61$.  This weaker bound would permit a massless
scalar.

As it is still possible that $\nu_\tau$ is massive and not relativistic
at the time of nucleosynthesis, we also show in Figure 1 the result of the
same calculation,
 when the probability distribution is allowed to go down to \N =
2 (i.e. the lower limit of integration in eq. (\ref{int}) is 2 rather than
3).  In this case, when
$\eta_{10} > 2.8$,
\N
$< 3.18$, a value closer to but still larger than the value of 3.13 found in
the
absence of renormalizing the distribution (even though the mean value is now
physical, a large portion of the Gaussian is still unphysical).
  Now, a massless scalar is clearly excluded.  It is interesting to note that,
 due to the renormalization procedure, the limit is actually weaker when three
 light or massless neutrinos are assumed.  This is due to the renormalization
 procedure.

In Figure 2, we show a similar plot with the
systematic errors described by a top-hat distribution. The convolved
Gaussian and top-hat is given by eq.(\ref{ytg}) where now $\mu$ corresponds
to the mean value of \N and the errors are the propagated errors in \N rather
than in
$Y_p$. The 95
\% CL upper limits are found by using the convolved distribution (\ref{ytg})
for $f(x)$ in eq. (\ref{int}). The limits are now somewhat tighter at low
$\eta_{10}$.  In this case for
$\eta_{10} > 2.8$, \N $< 2.91$ and 3.31 for \N $\ge 2$ and 3
respectively, excluding a massless scalar.

Our results clearly depend on the choice of systematic error.  In Figures 3
and 4, we show the sensitivity of our results to the
value of $\sigma_{\rm sys}$.  We plot
the 95 \% CL upper limit to \N as a function of  $\sigma_{\rm sys}$
for \N $\ge 2$ (solid curves) and \N $\ge 3$ (dashed curves) for three
choices of $\eta_{10} = 1.5$, $2.8$ and $4.0$.  The curves are arranged such
that the limits become more stringent (the upper limit on \N is lowered) as
$\eta_{10}$ is increased.  Figures 3 and 4 correspond to the double Gaussian
and Gaussian/top-hat approach described above.  Once again the latter
provides a more stringent upper bound.  As expected, as $\sigma_{\rm sys}$
is increased the limits on \N become significantly weaker.

In conclusion, we have shown that because the best fit values for the number
of neutrino-like particle species is, or is close to being unphysical, the
``true" upper limits on \N are significantly weaker than previous estimates.
The upper limit on \N is quite dependent on whether one
assumes \N $\ge2$ or 3.
 Nevertheless, big bang nucleosynthesis can and does supply us
with stringent constraint on \N and, more generally, on the speed-up of the
expansion rate of the
 Universe. For the perhaps more likely choice of \N $\ge 3$,
and for $\eta_{10}$ in
 the range 2.8 -- 4, the upper limit on \N ranges from 3.5 --
3.6 if systematic errors are treated as Gaussian and from 3.2 -- 3.3 if
systematic errors are treated as top-hats.  We stress that these
limits are only as good as the assumed errors (in particular the systematic
errors); the limit
 on \N is strongly dependent on $\sigma_{\rm sys}$. In the case
of Gaussian systematic
 errors, as  $\sigma_{\rm sys}$ approaches 0.01, the bound on
\N exceeds 4.0
 (if \N $\ge 3$).  If the systematic errors are treated as top-hats,
then the bound on
 \N is less sensitive and only exceeds 4.0 when $\sigma_{\rm sys}$
$ > .02$.  Finally we note that even if the observational errors could
eventually be
greatly diminished, there remains a residual theoretical error $\le
0.001$ in $Y$.  Because this translates into an error of about 0.08 in \N,
there
will always remain a residual uncertainty of order 0.08 in any determination of
\N.
Though we have argued here for somewhat weaker bounds on \N, they nevertheless
still provide us with a strong means for limiting particle physics beyond the
standard model.

\vskip 1.0truecm
\noindent {\bf Acknowledgments}
\vskip 1.0truecm
We especially thank Paul Langacker for calling to our attention the key issue
we
have discussed here and,
 Naoya Hata, Bob Scherrer, Dave Thomas and Terry Walker for
permission to utilize some of the results of our joint work in this manuscript.
We also would like to thank T. Falk and M. Roos for helpful conversations.
This work was supported in part by  DOE grants DE-FG02-94ER40823 and DE-AC02-
76ER01545.

\newpage
\noindent{\bf{Figure Captions}}

\vskip.3truein

\begin{itemize}

\begin{enumerate}

\item[{\bf Figure 1:}]  The 95 \% CL upper limit on \N as a function of
$\eta_{10}$ assuming that both statistical and systematic errors are
Gaussian distributed. The two curves correspond to the physical
condition that \N $>$ 2 or $>3$.  In the former case it must be assumed that
$\nu_\tau$ has a mass near its experimental upper limit.

\item[{\bf Figure 2:}] As in Figure 1, assuming that the
systematic errors are described by a top-hat distribution.

\item[{\bf Figure 3:}] The 95 \% CL upper limit on \N as a function of
the systematic uncertainty in $Y_p$. The solid (dashed) curves correspond to
the condition that \N $>$ 2 (3). Each of these two cases is shown for three
choices of $\eta_{10}$: 1.5, 2.8, and 4.0. The smaller values of $\eta$
yield weaker upper limits .

\item[{\bf Figure 4:}] As in Figure 3, assuming that the
systematic errors are described by a top-hat distribution.

\end{enumerate}
\end{itemize}


\begin{thebibliography}{99}
\bibitem{ssg} G. Steigman, D.N. Schramm, and J. Gunn,  Phys. Lett. {\bf B66}
(1977) 202.
\bibitem{wssok} T.P. Walker, G. Steigman, D.N. Schramm,
 K.A. Olive and K. Kang, Ap. J. {\bf 376}
 (1991) 51.
\bibitem{kern} D.Thomas, N. Hata, R.G. Scherrer, G. Steigman, G. \& Walker,
T.P. 1995,
(in preparation).
\bibitem{OSt} K.A. Olive and G. Steigman, Ap.J. Supp. {\bf 97} (1995) in press.
\bibitem{kk}P. Kernan and L.M. Krauss, Phys. Rev. Lett. {\bf 72}
(1994) 3309.
\bibitem{new} N. Hata, R. J. Scherrer, G. Steigman, D. Thomas,
T. P. Walker, S. Bludman and P. Langacker, (1995) in preparation.
\bibitem{rpp} Review of Particle Properties, Phys. Rev. {\bf D50}
(1994) 1173.
\bibitem{Kaw} E. W. Kolb and R. J. Scherrer, Phys. Rev {\bf D25} (1982) 1481 .
\bibitem{ytsso} J. Yang, M.S. Turner, G. Steigman, D.N. Schramm, and K.A.
Olive, Ap.J.  {\bf 281} (1984) 493.
\bibitem{vop} E. Vangioni-Flam, K.A. Olive, and N. Prantzos, Ap.J. {\bf
427} (1994) 618;
S. Scully and K.A. Olive, Ap.J. (in press) 1995.
\bibitem{orstv} K.A. Olive, R.T. Rood, D.N. Schramm, J.W. Truran, and E.
Vangioni-Flam, Ap.J. (in press) 1995;
M. Tosi, G. Steigman, and D.S.P. Dearborn, in
Proceedings of the ESO/EPIC Workshop on the Light Element Abundances, ed.
P. Crane (1994).
\bibitem{st} G. Steigman and M. Tosi, Ap.J. {\bf 401} (1992) 15;
OSU preprint OSU-TA-12/94 (1994).
\bibitem{hsstw} N. Hata, R.J. Scherrer, G. Steigman, D. Thomas, and T.P.
Walker, astro-ph/9412087, (1994).
\bibitem{cs}  R.F. Carswell, M. Rauch, R.J. Weymann, A.J. Cooke, and
J.K. Webb, MNRAS {\bf 268} (1994) L1; A. Songaila, L.L. Cowie,
C. Hogan, and M. Rugers,
Nature {\bf 368} (1994) 599.
\bibitem{s} G. Steigman, MNRAS, {\bf 269} L53.
\bibitem{vc} E. Vangioni-Flam and M.
Cass\'{e}, Ap.J. (in press) 1995.
\bibitem{t} D. Tytler, talk given at the Munich Absorption Line Meeting,
(1994).
\bibitem{sg} D. Sasselov and D. Goldwirth, preprint 1994.
\bibitem{cst} C.J. Copi, D.N. Schramm, and M.S. Turner, Science (in press)
1995.
\end{thebibliography}
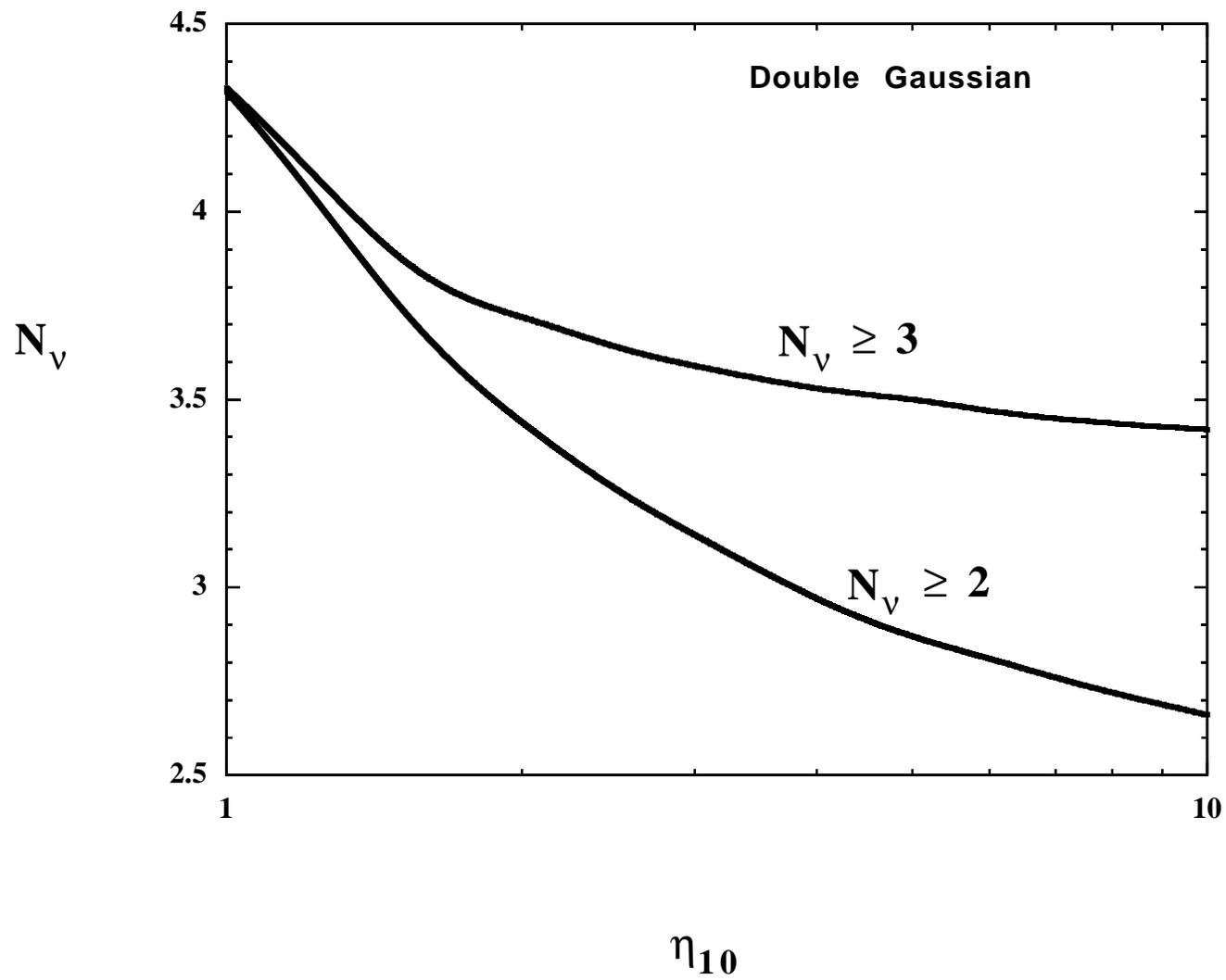
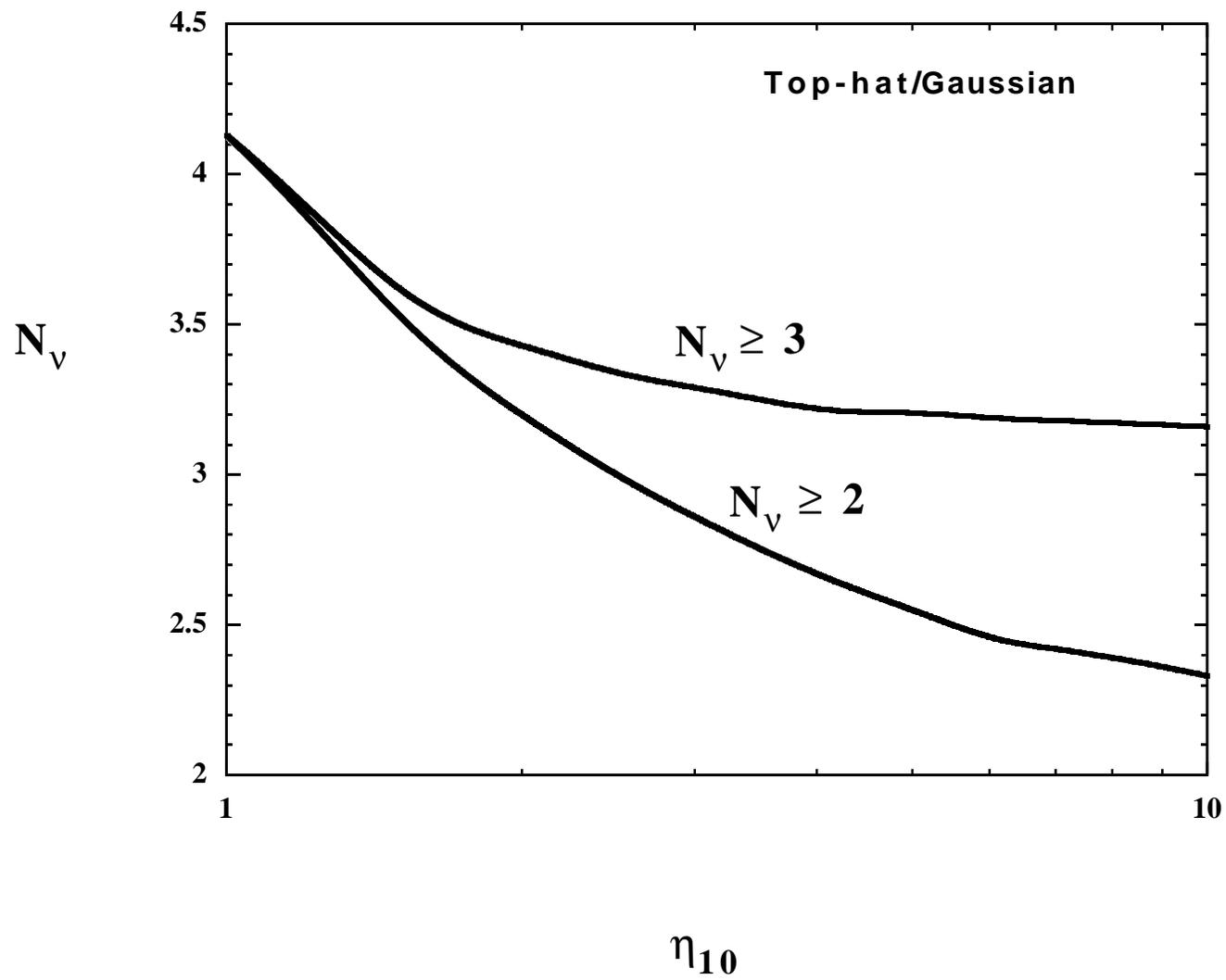
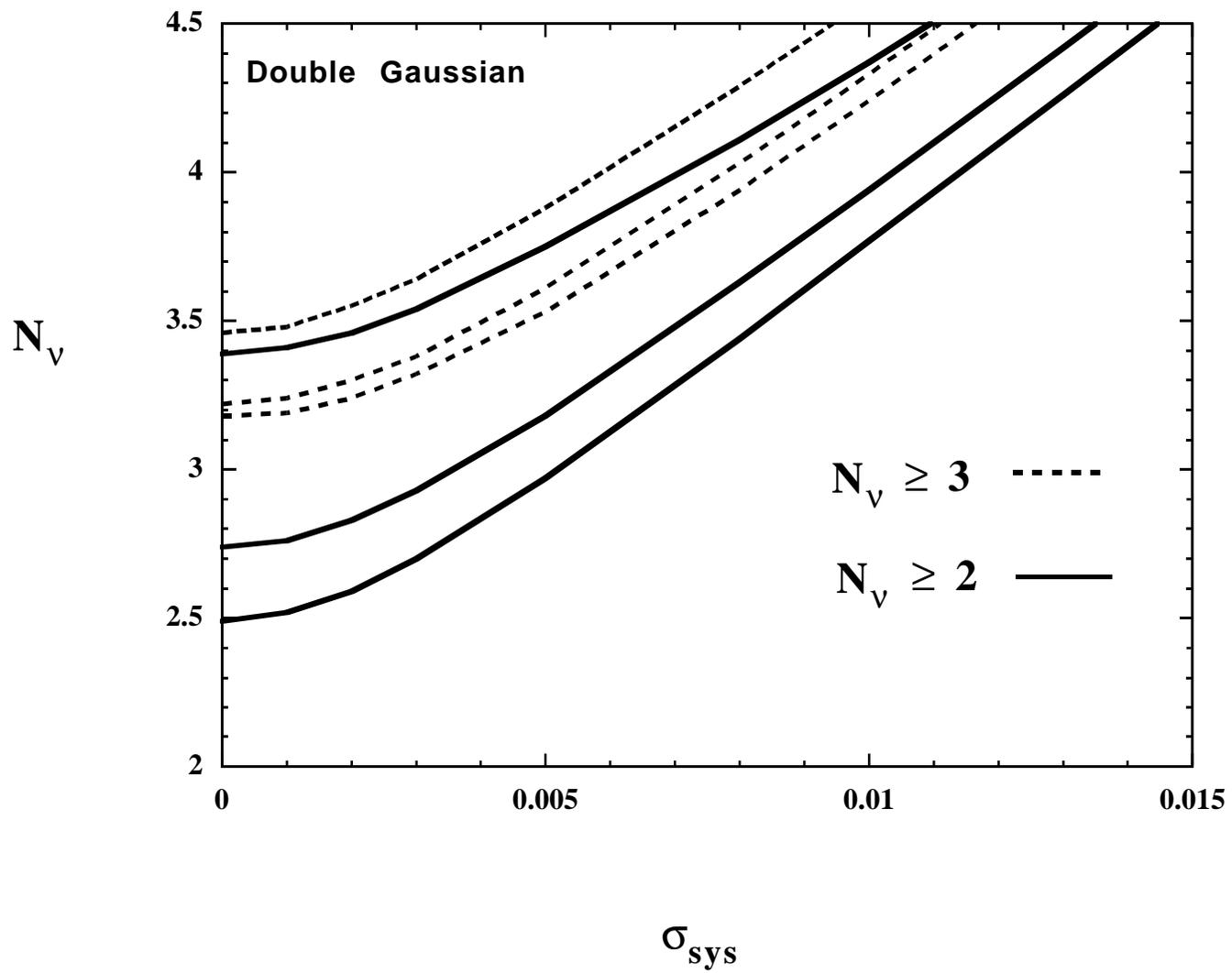
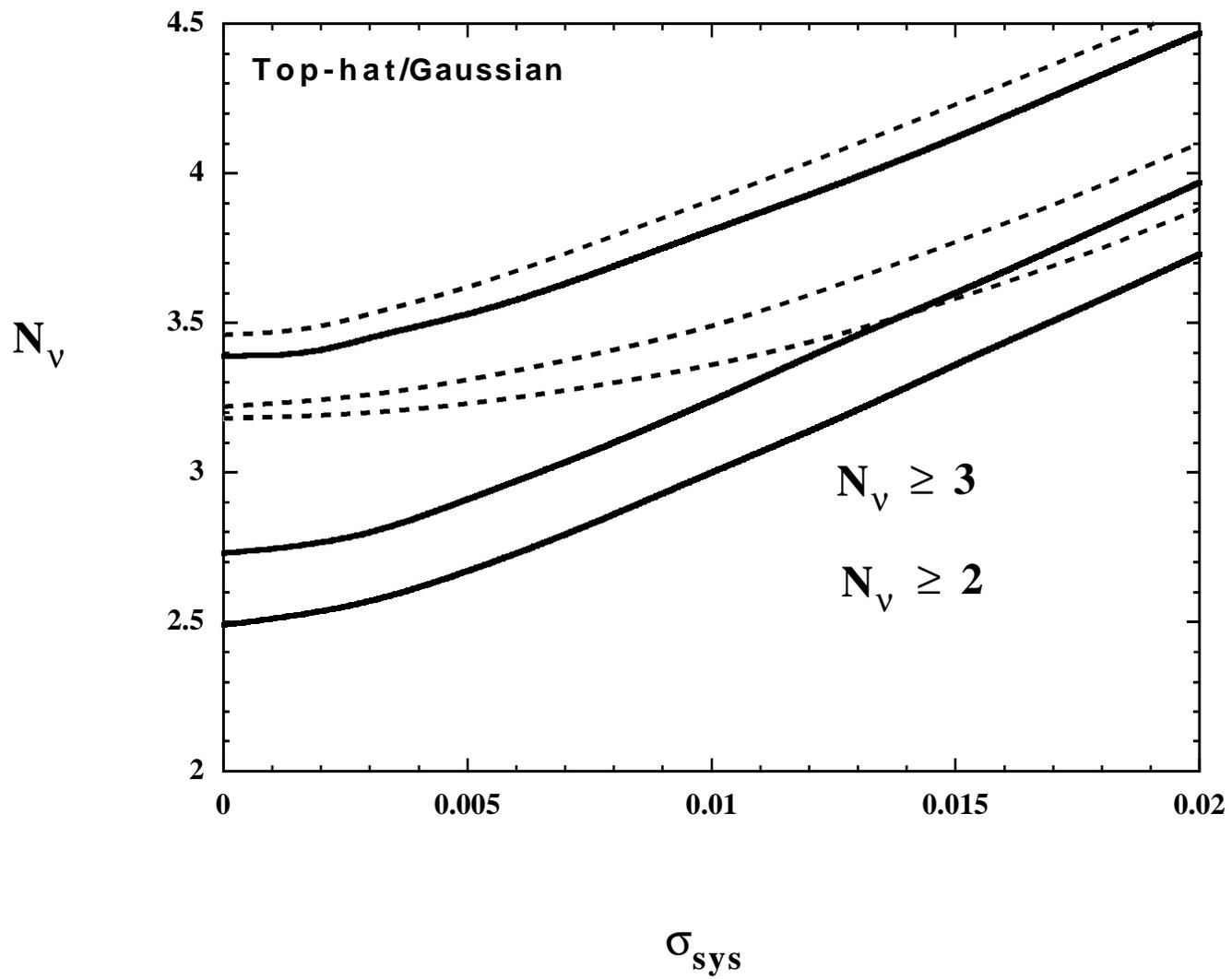
\end{document}